\documentclass[aps,superscriptaddress,showpacs,twocolumn,prr]{revtex4-2}

\usepackage[pdftex]{graphicx}
\usepackage{amssymb}
\usepackage{amsmath}
\usepackage{xspace}
\usepackage{color}
\usepackage{bm}
\usepackage{mathtools}
\usepackage{float}

\usepackage{hyperref}
\usepackage{natbib}

\usepackage{placeins}

\input{macro_f.lib}

\begin{document}

\title{Decay of spin helices in XXZ quantum spin chains with single-ion
anisotropy}

\author{Florian Lange}
\affiliation{Erlangen National High Performance Computing Center, Friedrich-Alexander-Universit\"at Erlangen-N\"urnberg, 91058 Erlangen, Germany}
\author{Frank G\"ohmann}
\affiliation{School of Mathematics and Natural Sciences, University of Wuppertal, 42097 Wuppertal, Germany}
\author{Gerhard Wellein}
\affiliation{Erlangen National High Performance Computing Center, Friedrich-Alexander-Universit\"at Erlangen-N\"urnberg, 91058 Erlangen, Germany}
\author{Holger Fehske}
\affiliation{Erlangen National High Performance Computing Center, Friedrich-Alexander-Universit\"at Erlangen-N\"urnberg, 91058 Erlangen, Germany}
\affiliation{Institute of Physics, University of Greifswald, 17489 Greifswald, Germany}

\begin{abstract}
	Long-lived spin-helix states facilitate the study of non-equilibrium dynamics in quantum magnets. We consider the decay of transverse spin-helices in antiferromagnetic spin-$S$ XXZ chains with single-ion anisostropy. The spin-helix decay is observable in the time evolution of the local magnetization that we calculate numerically for the system in the thermodynamic limit using infinite time-evolving block decimation simulations. Although the single-ion anisotropy prevents helix states from being eigenstates of the Hamiltonian, they still can be long-lived for appropriately chosen wave numbers. In case of an easy-axis exchange anisotropy the single-ion anisotropy may even stabilize the helices.  Within a spin-wave approximation, we obtain a condition giving an estimate for the most stable wave number $Q$ that agrees qualitatively with our numerical results.
\end{abstract}

\maketitle

\section{Introduction}
Spin helices are product states in one-dimensional spin systems that can be experimentally realized, observed, and manipulated \cite{PhysRevLett.113.147205,Jepsen2022,PhysRevX.11.041054}. They can remain stable when time evolved with the spin-$S$ Heisenberg XXZ Hamiltonian, as they are eigenstates, if a certain commensurability condition, called the phantom condition \cite{PhysRevB.104.L081410,Jepsen2022}, is satisfied. They may appear as reference states (or pseudo vacua) in exact Bethe ansatz calculations \cite{ZKP24,PhysRevLett.132.220404,PhysRevB.111.094437}. For wave vectors that dissatisfy the phantom condition only slightly, the helix states may still be long-lived \cite{Jepsen2022,PhysRevLett.132.220404}. They are robust against noise \cite{PhysRevB.107.214422}, and exist, in a more general form, in two and three dimensions as well \cite{PhysRevB.105.L060302,Jepsen2022,PhysRevResearch.7.L012008}. For these features, they play a prominent role in ongoing experimental and theoretical efforts to develop new and efficient quantum simulation techniques. In fact, long-time stability or slow dynamics of experimentally controllable initial states under quantum many-body time evolution, e.g., also in the context of many-body scars, are interesting in view of applications in future quantum technologies~\cite{Serbyn2021}, including quantum sensing~\cite{PRXQuantum.2.020330} and state transfer~\cite{StateTransferManyBodyScars}.

In this work we report our study of the spatio-temporal decay of spin helices in more general spin-$S$ chains time-evolved by the XXZ Hamiltonian with single-ion anisotropy, a case for which there are hardly any precise results available as yet, but which appears particularly interesting in connection with the equilibration dynamics in spin chains~\cite{SCMV24}. The stability of spin-helices is generally unrelated with integrability. Perfectly stable spin helices are supported by the non-integrable higher-spin XXZ chains \cite{Jepsen2022}. A single-ion anisotropy, however, destroys the perfect stability of all spin-helices and the underlying tower of scar states~\cite{6dbn-n6rv} in the higher-spin XXZ chains, while retaining a simple form of the corresponding time-evolved one-point functions. This should be contrasted with the effect of a uniform magnetic field in $z$ direction, which does not affect the decay of the spin helices in the XXZ model and only changes the phase velocity for the transverse spin component~\cite{Zhang_2024,PhysRevB.111.165106}.

As we shall see, the one-point functions of the XXZ chains with single-ion anisotropy form a helix that, at every moment of time, is determined by a single time-dependent expectation value of a vector operator. Its short-time behavior and overall instability can be accessed by Taylor-expanding the time-evolution operator. At longer times, compatible with the experimentally accessible scales, we have employed infinite time-evolving block decimation simulations to compute the relevant one-point functions numerically. Our precise numerical investigations reveal that single-ion anisotropy
causes the spin-helix amplitude to exhibit significantly more complex
and non-monotonous behaviour than in the simpler XXZ spin-chain model.
Surprisingly, the helix's lifetime can be considerable, even in the presence of strong single-ion anisotropy. We complement our study by comparing with conclusions drawn from an approximate spin-wave theory and find some of the numerically observed features at intermediate times at least qualitatively confirmed.

\section{Model and observables}
The antiferromagnetic XXZ chain with single-ion anisotropy is defined
by its Hamiltonian
\begin{equation}
    \hat{H} = J \sum_{n=1}^N \big(\hat{S}_n^x \hat{S}_{n+1}^x + \hat{S}_n^y \hat{S}_{n+1}^y + \Delta \hat{S}_n^z \hat{S}_{n+1}^z \big)
    + D \sum_{n=1}^N \big( \hat{S}_n^z \big)^2 .
	\label{eq:hamxxz}
\end{equation}
Here $N$ is the number of lattice sites, and periodic boundary conditions are implied. The operators $\hat{S}_n^{x,y,z}$ are spin-$S$ operators acting on site $n$ of the chain, $J > 0$ is the exchange energy, $\Delta > 0$ the exchange anisotropy, and the real parameter $D$ denotes the single-ion anisotropy. Throughout this work we set $J=1$.

The model has a rich ground-state phase diagram whose structure is not only determined by the model parameters but also depends significantly on the spin quantum number~$S$. Remarkably, for $S=1$, it even contains a symmetry-protected topological Haldane phase in addition to the large-$D$ and N\'eel phases~\cite{PhysRevLett.50.1153,LIU201463,10.21468/SciPostPhys.5.6.059,Ejima2021}.

We are interested in the time evolution of spin helices as reflected in the dynamics of the one-point functions of the local spin-$S$ operators. Let us define the total spin-$S$ operators $\hat{S}^\alpha = \sum_{n = 1}^N \hat{S}_n^\alpha$, $\alpha = x, y, z$, and a helix operator $\hat{\Phi} = \sum_{n = 1}^N (n-1) \hat{S}_n^z$. 
If $|{\uparrow}\>$ is the total spin highest-weight state, then the state
\begin{equation}
    |Q, \theta\> = \re^{- \i Q \hat{\Phi}} \re^{- \i \theta \hat{S}^y} |{\uparrow}\>
\end{equation}
represents a spin-helix of wave number $Q$  winding in the XY plane with polar angle $\theta$ against the $z$ axis. This type of state was experimentally realized for $S = 1/2$ in \cite{Jepsen2022}. We require $Q$ to be commensurate with the periodic boundary conditions, $QN\equiv 0$ $({\rm mod} \,2\p)$. 

Let $\hat{\Sv}_n = (\hat{S}^x_n, \hat{S}^y_n, \hat{S}^z_n)^t$ be the local spin-$S$ vector operator. The one-point functions that can be observed in the experimental setup of \cite{Jepsen2022} are $\<Q, \theta|\hat{\Sv}_n (t)|Q, \theta\>$, where $\hat{\Sv}_n (t)$ is the time-evolved spin operator in the Heisenberg picture. We shall use the notation $\ad_X$ for the adjoint Lie-algebra action of an Operator $X$, $\ad_ X Y = [X,Y]$, where $[.,.]$ is the commutator, and the notation $D_z (\ph)$ for the $3 \times 3$ matrix representing a rotation by $\ph$ about the $z$-axis in real space. Then $\re^{\i Q \ad_{\hat{\Phi}}} \hat{\Sv}_n = D_z (Qn) \hat{\Sv}_n$, and
\begin{align} \label{spacetimesep}
    \<Q, \theta|\hat{\Sv}_n (t)|Q, \theta\> &= \<0, \theta|\re^{\i Q \ad_{\hat{\Phi}}} \re^{\i t \ad_{\hat{H}}} \hat{\Sv}_n |0, \theta\> \nonumber \\
    &= D_z (Qn) \<0, \theta| \re^{\i t \ad_{\hat{H}_Q}} \hat{\Sv}_1|0, \theta\>,
\end{align}
where
\begin{multline}
    \hat{H}_Q = \re^{\i Q \ad_{\hat{\Phi}}} \hat{H} =
        - \sin(Q) \sum_{n=1}^N \bigl(\hat{S}_n^x \hat{S}_{n+1}^y - \hat{S}_n^y \hat{S}_{n+1}^x \bigr) \\
        +  \sum_{n=1}^N \Bigl[\cos(Q) \bigl(\hat{S}_n^x \hat{S}_{n+1}^x + \hat{S}_n^y \hat{S}_{n+1}^y \bigr)
          + \Delta \hat{S}_n^z \hat{S}_{n+1}^z + D \big( \hat{S}_n^z \big)^2\Bigr]
          \label{eq:ham_transformed}
\end{multline}
and where we have used the translation invariance of $\hat{H}_Q$ in the second equation in (\ref{spacetimesep}).

Equation~(\ref{spacetimesep}) shows that the spatial degrees of freedom in the time evolution of the spin helix are entirely determined by the factor $D_z (Qn)$. At every fixed moment of time the one-point functions form a helix in space. The temporal behavior is determined by a single vector-valued quantity $\<0, \theta| \re^{\i t \ad_{\hat{H}_Q}} \hat{\Sv}_1|0, \theta\>$. In this sense, spatial and temporal degrees of freedom are decoupled (see \cite{PhysRevB.107.235408}). It is therefore natural to study the one-point function
\begin{equation}
    \Sv_{Q, \theta} (t|\D, D, S) = \lim_{N \rightarrow \infty} \frac{\<0, \theta| \re^{\i t \ad_{\hat{H}_Q}} \hat{\Sv}_1|0, \theta\>}{S \sin(\theta)}.
    \label{eq:onepoint}
\end{equation}
The case $D = 0, \ S = 1/2$ was considered in \cite{PhysRevB.107.235408} based on a short-time expansion and numerical analysis. If in addition $\theta = \p/2, \D = 0$, the function $\Sv_{Q, \theta}$ and its long-time asymptotic behavior can be calculated exactly \cite{PhysRevLett.132.220404}.

Certain symmetries worked out for $S = 1/2$ in \cite{PhysRevB.107.235408} persist in our more general case. In particular, using that $|0, \theta\>$ is invariant under spatial reflections, while $\hat{H}_Q \mapsto \hat{H}_{- Q}$, we infer that $\Sv_{Q, \theta} (t|\D, D, S) = \Sv_{- Q, \theta} (t|\D, D, S)$. Performing a rotation by $\p$ about the $x$ axis, we see that $\re^{- \i \p \hat{S}^x} |0, \p/2\> = |0, \p/2\>$ and $\re^{\i \p \ad_{\hat{S}^x}} \hat{\Sv}_n = \diag(1,-1,-1) \hat{\Sv}_n$. Hence,
\begin{equation}
    \Sv_{Q, \p/2} (t|\D, D, S) \\ = \diag(1,-1,-1) \Sv_{Q, \p/2} (t|\D, D, S),
\end{equation}
implying that $S_{Q, \p/2}^y = S_{Q,\p/2}^z = 0$. Thus, if the helix is initially in the XY plane, a single scalar function 
$S_{Q, \p/2}^x (t|\D, D, S)$ determines its time dependence. If this is not the case, two functions are necessary in order to characterize the time dependence of the transverse components of the helix. We may, for instance, take $S_{Q, \theta}^x$ and $S_{Q,\theta}^y$ or, alternatively, an amplitude $A$ and a phase function $\phi$:
\begin{equation} \label{aandphi}
	A = \frac{1}{S} \sqrt{\bigl(S_{Q, \theta}^x\bigr)^2 + \bigl(S_{Q, \theta}^y\bigr)^2}, \ \
    \tan(\phi) = - S_{Q,\theta}^y/S_{Q,\theta}^x.
\end{equation}
The amplitude $A$ is equivalent to the spin contrast that was measured for spin-1/2 XXZ chains in optical lattice systems~\cite{PhysRevLett.113.147205,Jepsen2022,PhysRevX.11.041054}. 

\section{Spin-helix amplitude decay}
\begin{figure}[tb]
\centering
	\includegraphics[width=0.95\linewidth]{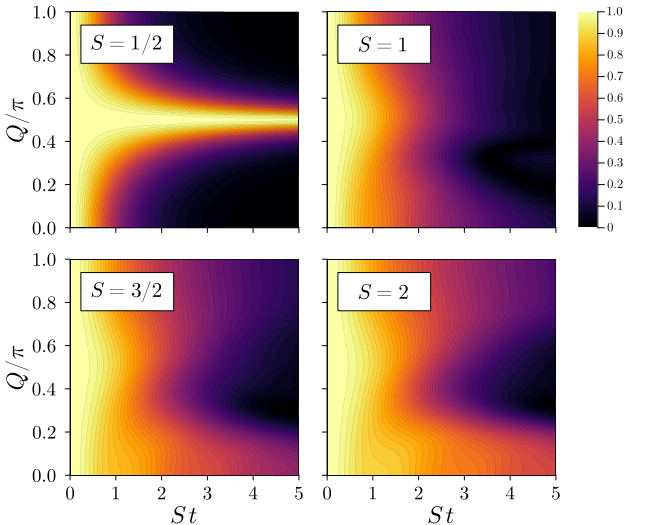}
	\caption{Contour plot showing the spin-helix amplitude $A$ as a function of wave number $Q$ and rescaled time $S t$ for $\Delta = 0$, $D=0.5$ and $\theta=\pi/2$. For $S=1/2$, the helix with $Q/\pi = 0.5$ is an eigenstate and therefore $A$ does not decay. In contrast, no stable helix is observed for $S > 1/2$.} 
	\label{fig:color_Delta0}
\end{figure}
For short times we can expand the time evolution operator on the right hand side of Eq.~\eqref{eq:onepoint} in a power series in $t$. Using the properties of spin-$S$ operators under complex conjugation we see that $\Sv_{Q, \theta} (- t|\Delta, D, S) = \diag(1,-1,1) \Sv_{Q, \theta} (t|\Delta, D, S)$ and therefore $A(-t) = A(t)$ and $\phi(-t) = - \phi(t)$. Hence, the short-time expansion gives $A$ as a series in even powers of $t$. For $\theta = \pi/2$ we obtain
\begin{equation}
	A(t) =  1 - \frac{t^2}{2} \left[(2S - 1)D^2 + S \big( \Delta - \cos(Q)\big)^2 \right]
	  + \mathcal{O}(t^4) .
	\label{eq:decay_2ndO}
\end{equation}
This shows that a necessary condition for a non-decaying spin helix is $S =1/2$ or $D = 0$, and
\begin{equation} \label{phantomas}
    \cos(Q) = \Delta .
\end{equation}
The latter condition is the ``phantom condition'' of \cite{PhysRevB.104.L081410}. In \cite{PhysRevB.104.L081410} it was shown that this condition guaranties the existence of stable helices if $S = 1/2$. In general, we see that the larger $S$ or $D$, the faster is the initial decay of the helices. For fixed $S$ and $D$ the initial decay is slowest, if the phantom condition Eq.~\eqref{phantomas} is satisfied.

In order to hold control over the time evolution at intermediate time-scales we use the infinite time-evolving block decimation (iTEBD) algorithm~\cite{iTEBD} with a second-order Suzuki-Trotter decomposition to simulate the time evolution of the spin helix in the limit of an infinite system size $N \rightarrow \infty$ numerically. We employ the representation in Eq.~\eqref{eq:onepoint}, since the translation invariance of $\hat{H}_Q$ facilitates the use of the regular iTEBD with unit cell 2. The main error sources are the Suzuki-Trotter discretization and the truncation of the bond dimension in the infinite matrix-product state~\cite{SchollwoeckMPSReview}. We found a time step of $0.02/S$ and a maximum bond dimension of 3000 to give accurate results over the considered time ranges, with the truncation error after each gate remaining below $5\cdot 10^{-8}$.

\begin{figure}[tb]
\centering
	\includegraphics[width=0.95\linewidth]{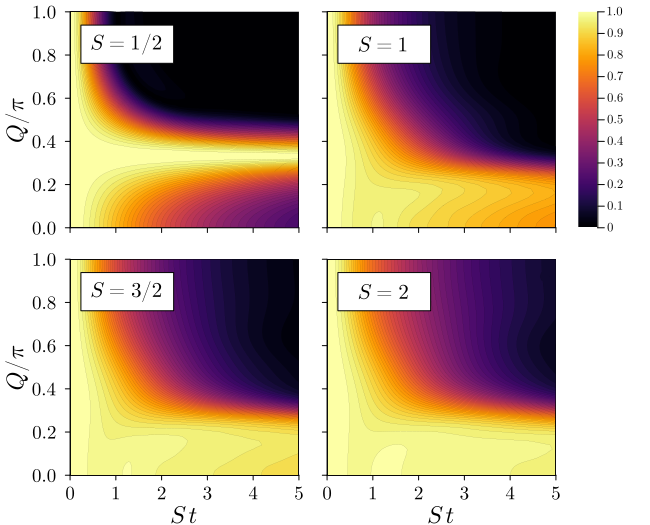}
	\caption{Same as Fig.~\ref{fig:color_Delta0} but for $\Delta = 0.5$, $D=0.5$ and $\theta=\pi/2$. The spin-helix amplitude decays noticeably slower for a range of $Q$ values, but the most stable point $\tilde{Q}$ for $S> 1/2$ is shifted compared to the phantom condition in the XXZ chain.} 
	\label{fig:color_Delta0.5}
\end{figure}

\begin{figure}[tb]
\centering
	\includegraphics[width=0.99\linewidth]{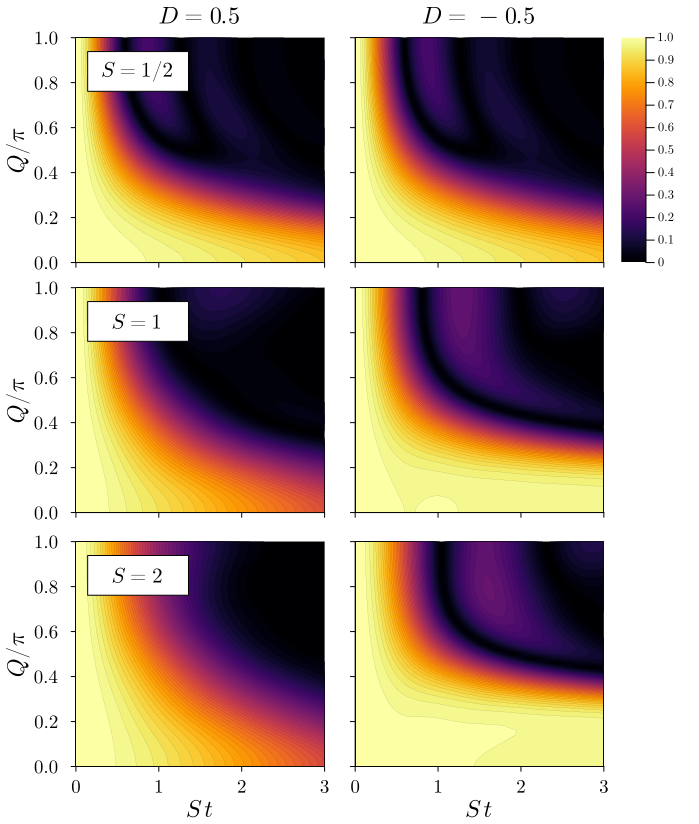}
	\caption{Spin-helix amplitude for $\Delta = 1.2$, $D = \pm0.5$ and $\theta=\pi/2$. A negative single-ion anisotropy $D=-0.5$ leads to slowly decaying helix states even in the easy-axis regime. } 
	\label{fig:color_Delta1.2}
\end{figure}

We first discuss our numerical results for the simpler case of spin helices with polar angle $\theta = \pi/2$, when the one-point functions are fully determined by the amplitude $A$. Figures~\ref{fig:color_Delta0}, \ref{fig:color_Delta0.5} and \ref{fig:color_Delta1.2} show the amplitude $A$ as a function of the wave number $Q$ for $1/2 \leq S \leq 2$. Time $t$ is scaled with $S$ in view of the classical limit $S \rightarrow \infty$. For $S=1/2$ we observe stable helices once the phantom condition Eq.~\eqref{phantomas} is satisfied (Figs.~\ref{fig:color_Delta0} and \ref{fig:color_Delta0.5}). Note that in the special case $S=1/2$ and $\Delta = 0$, our numerical results agree with those of the exact analytical expression~\cite{PhysRevLett.132.220404} (see the Appendix). 
For larger values of $S$ the single-ion anisotropy leads to a fast initial decrease of $A$ at times $S t \lesssim 1$ for all wave numbers $Q$, in accordance with Eq.~(\ref{eq:decay_2ndO}).

At longer times we see a remnant of the stable spin helix, at least for $\Delta = 0.5$, where the decay is noticeably slower after the initial transient period for a small range of wave numbers $Q$. We define $\tilde{Q}$ to be the wave number for which the spin-helix amplitude shows the slowest decay. A single-ion anisotropy with $D=0.5$ shifts $\tilde{Q}$ to smaller values compared to the phantom condition Eq.~\eqref{phantomas}, valid at $D=0$. As demonstrated in Fig.~\ref{fig:color_Delta1.2} for $\Delta = 1.2$, a negative $D$ has the opposite effect. It moves $\tilde{Q}$ to larger values and can even stabilize a spin helix in the easy-axis regime. The results for $\Delta = 0$ and $D=0.5$ in Fig.~\ref{fig:color_Delta0} do not seem to fit into this simple picture, as the spin helices decay relatively quickly for all $Q$ and $1 \leq S \leq 2$ in this case.  

\begin{figure}[tb]
\centering
        \includegraphics[width=0.825\linewidth]{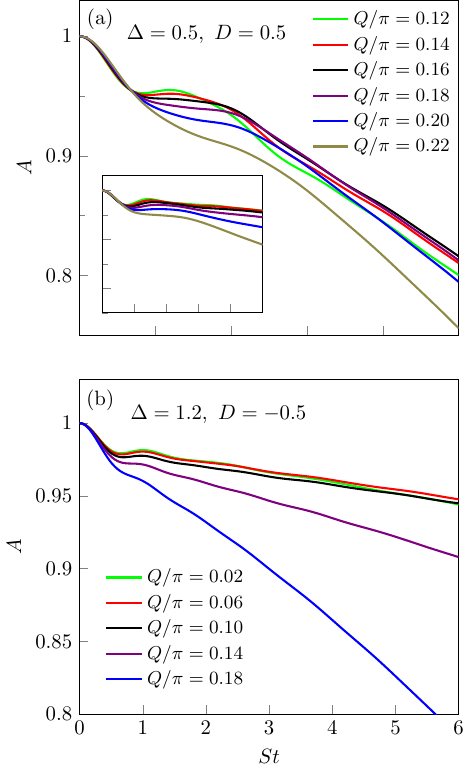}
	\caption{Time evolution of the spin-helix amplitude for $S=1$ and $\theta = \pi/2$ near the wave number $Q$ with the slowest decay. Panel (a) and (b) are for parameters $(\Delta,D)=(0.5,0.5)$ and $(\Delta,D)=(1.2,-0.5)$, respectively. The inset in (a) displays $A(t)$ for $S=2$ using the same range for the axes.}
        \label{fig:decay_closeup}
\end{figure}

The time dependence of the amplitude $A$ for $Q$ close to $\tilde{Q}$ and $\Delta = 0.5$ is shown in more detail in Fig.~\ref{fig:decay_closeup}(a). After an initial quadratic decrease in accordance with Eq.~\eqref{eq:decay_2ndO}, the amplitude $A(t)$ stabilizes for a short time before it decreases further with almost constant slope. This is quite different from the model without single-ion anisotropy, which does not exhibit a slowdown of the decay at intermediate times~\cite{PhysRevB.107.235408,PhysRevLett.132.220404,XYZHelicesScars}. Increasing the spin from $S=1$ to $S=2$ induces a considerably slower decrease at long times and shifts the wave number of slowest decay $\tilde{Q}$ toward smaller values. In the classical limit $S \rightarrow \infty$ the spin helix becomes stable for all $Q$, and $A(t) = 1$.

\begin{figure}[tb]
\centering
        \includegraphics[width=0.825\linewidth]{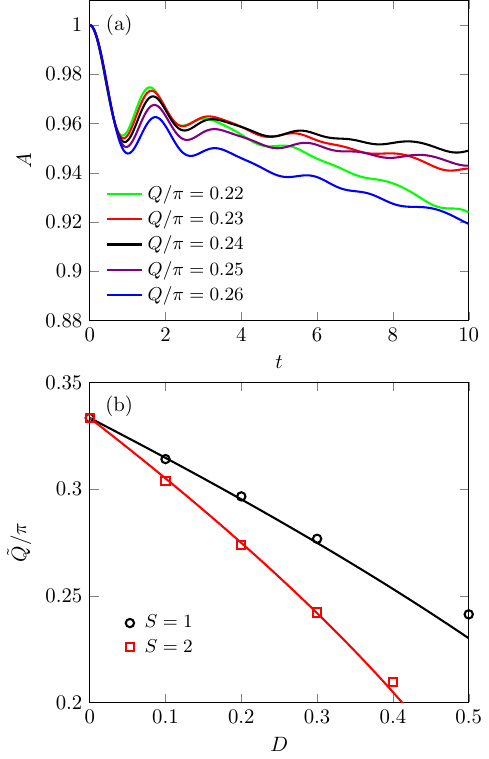}
	\caption{(a) Spin-helix amplitude in the model Eq.~\eqref{eq:ham_transformed} without Dzyaloshinskii-Moriya interaction and parameters $\Delta=0.5$, $D=0.5$ and $\theta=\pi/2$ [panel (a)]. In (b), the wave number $\tilde{Q}$ for which the spin helix decays the slowest is compared with the prediction Eq.~\eqref{eq:stable_condition} (solid lines). Here, we defined $\tilde{Q} = \text{arg max}_Q \int_{8/S}^{10/S} A(t) \, dt$.} 
        \label{fig:Qshift}
\end{figure}

The spin helices appear to be more stable for wave numbers $\tilde{Q}$ near zero. This is shown in Fig.~\ref{fig:decay_closeup}(b) for model parameters $\Delta = 1.2$ and $D=-0.5$, corresponding to a minimal decay at $Q / \pi \simeq 0.06 $. A possible explanation is that the long-time decrease of the amplitude is amplified by the Dzyaloshinskii-Moriya interaction in $\hat{H}_Q$, Eq.~\eqref{eq:ham_transformed}, which scales with $\sin(Q)$ and is therefore suppressed for long wavelengths.
This is supported in Sec.~\ref{sec:spinwave} by a spin-wave model and the numerical results in Fig.~\ref{fig:Qshift}. 

\begin{figure}[tb]
\centering
        \includegraphics[width=0.9\linewidth]{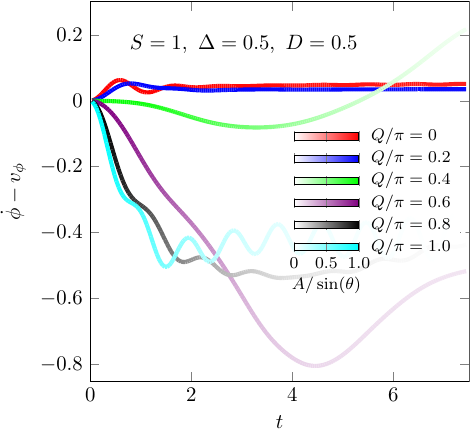}
	\caption{Difference between the phase velocity $\dot{\phi}(t)$ and the semiclassical prediction $v_\phi$ given by Eq.~\eqref{eq:SCS}. The parameters are $S=1$, $\Delta = 0.5$, $D=0.5$ and $\theta=\pi/4$.} 
	\label{fig:phase_velocities}
\end{figure}

\section{Phase velocity}
So far we have discussed planar spin helices with $\theta = \pi/2$ for which the phase $\phi$ remains zero during the time evolution. As an example of a non-planar helix displaying a non-trivial phase dynamics, we now consider the case $\theta = \pi/4$. Employing again the short-time expansion of $\Sv_{Q,\theta}$, taking into account that $S_{Q, \theta}^x$ is even as a function of $t$, while $S_{Q, \theta}^y$ is odd, we obtain from Eq.~\eqref{aandphi} that $\phi (t) = v_\phi t + {\cal O} (t^3)$, where 
\begin{equation}
    v_\phi = 2 S \cos(\theta) \bigl(\cos(Q) - \Delta - D [1 - 1/(2S)]\bigr)  .
    \label{eq:SCS}
\end{equation}

Remarkably, the semiclassical equations of motion \cite{Schliemann_1998} imply a phase velocity which is equal to $v_\phi$. For $S = 1/2$ and $0 \le \Delta < 1$ it was shown \cite{PhysRevB.107.235408} that the phase velocity $\dot \phi (t) = \frac{\rd}{\rd t} \phi (t)$ rapidly approaches an asymptotic value that generically differs from the initial value $\dot \phi (0) = v_\phi$. Figure~\ref{fig:phase_velocities} shows the difference $\dot \phi (t) - v_\phi$ as a function of time for $\Delta=0.5$, $D=0.5$ and $S=1$. In the region of small wave numbers $Q$, where the spin helix is relatively stable, we observe that $\phi$ increases at short times, but then stays nearly constant. For larger $Q$, the phase velocity changes more rapidly with time and does not seem to converge before the amplitude $A$ of the spin-helix dwindles.

The spin helix $|Q,\theta \rangle$ is a superposition of states in a tower $| S_\ell \rangle = \big( \hat{S}_Q^+ \big)^\ell | {\downarrow} \rangle$ ($\ell \in \{ 0, 1,...,2SN\}$) that is created by the repeated application of a magnon operator $\hat{S}_Q^+ = \sum_n e^{- \textrm{i} Q n} \hat{S}_n^+$, starting from the total spin-$S$ lowest-weight state $| {\downarrow} \rangle$~\cite{PhysRevB.111.165106, 6dbn-n6rv}. For the transverse spin component, the relevant matrix elements are of the form $\langle S_{\ell + 1} | \hat{S}_Q^+ (t) | S_{\ell} \rangle \propto e^{\textrm{i}t(E_{\ell + 1} - E_{\ell}) } + \mathcal{O}(t^2)$, where $E_{\ell}=\langle S_{\ell} |\hat{H} | S_{\ell} \rangle / \langle S_{\ell} | S_{\ell}\rangle$, which shows that the phase velocity at short times is determined by the energy difference between states separated by one magnon. The weight $|\langle S_{\ell} | Q, \theta \rangle |$ is concentrated around basis states with $\ell /N = S [1 - \cos(\theta)]$, for which the energy difference is given by the right-hand side of Eq.~\eqref{eq:SCS}.

\section{Spin-wave approximation} \label{sec:spinwave}
The stability of spin helices in spin-$S$ chains without single-ion anisotropy was recently analyzed using a spin-wave approximation~\cite{XYZHelicesScars}. A particular result was that the decay is not symmetric in deviations from the stable wave number $Q$. 
The approach may be generalized to include finite $D$. In that case the construction of a suitable spin-wave Hamiltonian becomes more subtle, since the usual procedure of truncating a Holstein-Primakoff transformation and keeping the quadratic part after normal ordering leads to $D$-dependent expressions even for $S=1/2$. We consider planar helices and the following approximation 
to $\hat{H}_Q$ \cite{Tsuru1986}: 
\begin{align}
    \frac{\hat{H}_{SW}}{S} &= \sum_k \mathcal{A}_k \hat{a}_k^\dagger \hat{a}_k + \sum_k \frac{\mathcal{B}_k}{2} \left( \hat{a}_k \hat{a}_{-k} + \hat{a}_k^\dagger \hat{a}_{-k}^\dagger  \right) 
	\label{eq:sw}, 
\end{align}
where $\mathcal{A}_k = [ \cos(Q) + \Delta] \cos(k) - 2 \cos(Q) + D[1 - 1/(2S)]$, $\mathcal{B}_k = [\cos(Q) - \Delta] \cos(k) - D[1 - 1/(2S)] $, and $\hat{a}_k^{( \dagger)}$ are bosonic annihilation (creation) operators. Note that this spin-wave model is based on the state polarized in the $x$ direction. 

For a periodic $N$-site system the Bose operators are related to the spin-helix amplitude by $A = 1 - \sum_k \hat{a}_k^\dagger \hat{a}_k /(N S) $. 
Applying the result of Ref.~\cite{XYZHelicesScars} to the above model yields the condition 
\begin{align}
	\cos(Q) - \Delta - D[1 - 1/(2S)] = 0
	\label{eq:stable_condition}
\end{align}
for a constant amplitude, which is equal to the requirement for the semiclassical phase velocity $v_\phi$ to become zero. 

Our numerical results show that Eq.~\eqref{eq:stable_condition} does not imply that the exact amplitude is constant. At finite $D$ it rather continues to decrease with time for all $Q$. The wave number $\tilde{Q}$ with the smallest decay does not match Eq.~\eqref{eq:stable_condition} either [see Fig.~\ref{fig:decay_closeup}(a)]. However, 
it seems to capture the dependence of $\tilde{Q}$ on $D$ and $S$ at least qualitatively. The leading terms that were dropped in Eq.~\eqref{eq:sw} behave as $S^{-\frac{1}{2}}$ and correspond to the Dzyaloshinskii-Moriya interaction in Eq.~\eqref{eq:ham_transformed}. This interaction does not contribute to the second order in $t$ [see Eq.~\eqref{eq:decay_2ndO}] but strongly affects the long-time behavior. In fact, when those terms are dropped in the Hamiltonian~\eqref{eq:ham_transformed}, Eq.~\eqref{eq:stable_condition} accurately predicts $\tilde{Q}$, and the long-time decay of the amplitude $A$ for that wave number is severely reduced. This is demonstrated in Fig.~\ref{fig:Qshift} for an exchange anisotropy $\Delta = 0.5$. 
We note that removing the Dzyaloshinskii-Moriya interaction from the transformed Hamiltonian corresponds to adding a term 
$\delta \hat{H} = \frac{\textrm{i} J \sin(Q)}{2} \sum_n (e^{\textrm{i}Q} \hat{S}_n^+ \hat{S}_{n+1}^- - e^{-\textrm{i}Q} \hat{S}_n^- \hat{S}_{n+1}^+)$
to the original Hamiltonian which then is a special case of the model studied in Ref.~\cite{PhysRevB.108.085108}. 
With this term, already for $D=0$ the helices with opposite winding direction, $|{-}Q,\theta\rangle$, are no longer eigenstates of the Hamiltonian and the degeneracy at the helix energy is reduced.

When Eq.~\eqref{eq:stable_condition} holds, the spin-wave approximation for the amplitude decay can be written in closed form: 
\begin{align}
A(t) = 1 - \frac{1}{8S} \frac{[\Delta -  \cos(Q)]^2}{ \Delta |\cos(Q)|} \left[1 - \cos\left(\frac{\omega t}{2} \right) J_0\left(\frac{\omega t}{2} \right) \right], \label{eq:ASW}
\end{align} 
where $\omega = 8 S \sqrt{ \Delta |\cos(Q)|}$, and $J_0(.)$ is a Bessel function of the first kind. Although this is not quantitatively correct for small $S$, as can already be recognized by comparing the short-time expansion with Eq.~\eqref{eq:decay_2ndO}, the appearance of oscillations and the tendency of a stronger decay for small $\Delta$ or $|\cos(Q)|$ can also be seen in the numerical results (see the Appendix for a direct comparison). In the full model, the oscillations are much more dampened (Fig.~\ref{fig:decay_closeup}), however. We think that the algebraic decay to a constant amplitude that follows from the above expression for $A(t)$ is likely to be an artifact of the spin-wave approximation.

\section{Conclusions}
We have studied numerically how a single-ion anisotropy affects the decay of transverse spin helices in a spin-$S$ XXZ chain. Compared to the XXZ spin-$1/2$ model, in which a single-ion anisotropy cannot be effective, we found a more complex non-monotonic behavior of the spin-helix amplitude with an initial quadratic decay at short times and a subsequent decrease that strongly depends on the wave number $Q$ of the helix. Interestingly, by tuning $Q$, the spin helix can be made long-lived even for large $D$, if the exchange anisotropy $\Delta$ does not deviate too much from one. Within a spin-wave approximation~\cite{XYZHelicesScars}, requiring a non-decaying amplitude results in a simple expression for the wave vector $Q$ in terms of $D$ and $S$. This relation may be considered as a generalization of the phantom condition. However, while the phantom helices are eigenstates of the spin-$1/2$ XXZ chain and are connected with exact quantum many-body scars, the helices considered here slowly decay with time as can be seen from our numerical analysis. This is in line with recent work on perturbed many-body scars, where slow thermalization was observed and a lower bound for the thermalization time was derived~\cite{PhysRevResearch.2.033044}. What we find remarkable is that the spin-helices can be long-lived even for large $D$.

A question for future study would be how generalizations of spin helices in higher-dimensional lattices~\cite{Jepsen2022,PhysRevB.105.L060302} or other graphs~\cite{PhysRevResearch.7.L012008} decay in the presence of a single-ion anisotropy. Long-lived helices are likely to exist in such systems as well, but the conditions on $D$ and the spin angles will be different. For the square lattice, the spin-wave approximation suggests a stable helix for $\Delta + D[1 - 1/(2S)]/2 - \cos(Q_{x,y}) = 0$. It is not obvious, however, how the dynamics deviate in the full model and whether there are qualitative differences compared to the one-dimensional case.

It was recently demonstrated that other spin models~\cite{wang2024generalizedspinhelixstates}, including the XYZ chain~\cite{6dbn-n6rv,bhowmick2025granovskiizhedanovscarxyzspinchain}, support scar states in the form of generalized spin helices with more complex spin profiles, and that they relax much slower under some perturbations~\cite{XYZHelicesScars} than expected based on the general bounds given in Ref.~\cite{PhysRevResearch.2.033044}. 
The time dependence of the helix amplitude in these cases differs from the decay due to the single-ion anisotropy considered here, however. Understanding the relaxation dynamics and establishing whether thermalization occurs for spin helices under typical perturbations would be important goals for further research. \\

\section*{acknowledgments}
The authors gratefully acknowledge the scientific support and HPC resources provided by the Erlangen National High
Performance Computing Center (NHR@FAU) of the Friedrich-Alexander-Universit\"at Erlangen-N\"urnberg (FAU). NHR funding is provided by federal and Bavarian state authorities. NHR@FAU hardware is partially funded by the German Research Foundation (DFG) – 440719683. 

MPS simulations were performed using the ITensor library~\cite{itensor,itensor-r0.3}.

\appendix
\setcounter{equation}{0}\renewcommand\theequation{A\arabic{equation}}
\section*{Appendix} \label{sec:appendix}
\begin{figure}[tb]
\centering
        \includegraphics[width=0.95\linewidth]{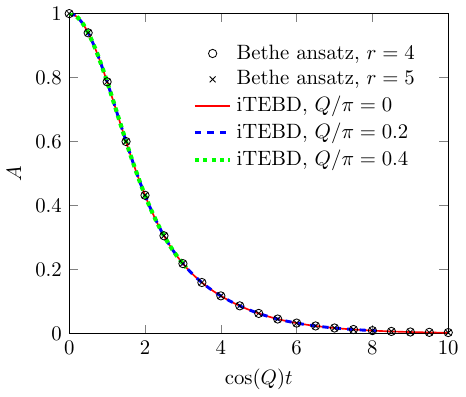}
	\caption{Comparison of the Bethe ansatz expression Eq.~\eqref{eq:ABethe} and iTEBD results for $S=1/2$, $\Delta = 0$ and $\theta= \pi / 2$.}
    \label{fig:Bethe}
\end{figure}

For $S=1/2$, $\Delta=0$ and $\theta = \pi / 2$, an exact expression for the time evolution of the spin-helix amplitude $A$ has been derived using Bethe ansatz techniques~\cite{PhysRevLett.132.220404}: 
\begin{align}
A(t_Q) &= \lim_{r\rightarrow \infty} A(r,t_Q), \label{eq:ABethe} \\
A(r,t_Q) &= \left| \det_{m,n=1,...,r} B_{m,n}(t_Q) \right|^2, \label{eq:ABethe_r} \\
B_{m,n}(t_Q) &= \delta_{m,n} + K_{m,n}(t_Q) + K_{m,1-n}(t_Q),
\end{align}
where $t_Q = \cos(Q) t$ and 
\begin{widetext}
\begin{align}
  K_{m,n}(t_Q) = &\frac{t_Q}{m-n} \left[ J_{2m}(t_Q) J_{2n-1}(t_Q) - J_{2n}(t_Q)J_{2m-1}(t_Q)\right] + \frac{t_Q}{m-n} \left[ J_{2m-1}(t_Q) J_{2n-2}(t_Q) - J_{2n-1}(t_Q)J_{2m-2}(t_Q)\right] \nonumber \\
  & + \frac{\textrm{i}t_Q}{m-n-1/2} \left[ J_{2m-2}(t_Q) J_{2n}(t_Q) - J_{2n-1}(t_Q)J_{2m-1}(t_Q) \right] \nonumber \\ &- \frac{\textrm{i}t_Q}{m-n+1/2} \left[ J_{2m-1}(t_Q) J_{2n-1}(t_Q) - J_{2n-2}(t_Q)J_{2m}(t_Q) \right]. \label{eq:Kmn}
\end{align}
\end{widetext}
In Eq.~\eqref{eq:Kmn}, $J_k(.)$ are Bessel functions and the diagonal elements should be understood as limiting values $K_{n,n} = \lim_{m\to n}K_{n,m}$. Small values of $r$ in Eq.~\eqref{eq:ABethe} already give highly accurate approximations for the decay of the amplitude $A$. 
As shown in Fig.~\ref{fig:Bethe}, there is no appreciable difference between $r=4$, $r=5$ and the iTEBD results for the time range considered in this work.

\begin{figure}[tb]
\centering
        \includegraphics[width=0.95\linewidth]{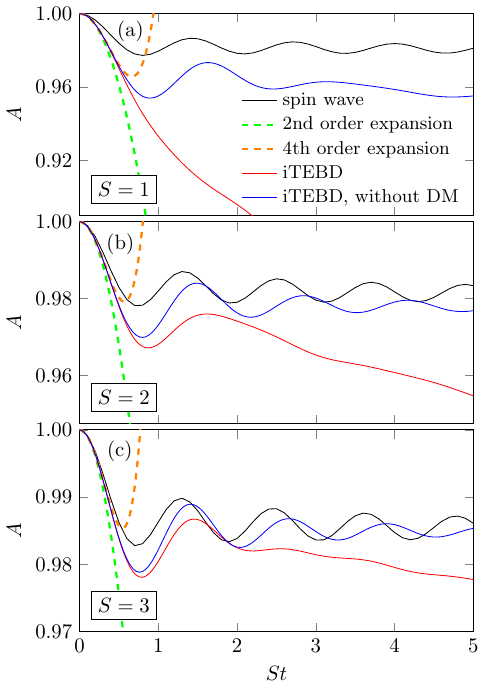}
	\caption{Comparison of iTEBD results [with and without Dzyaloshinskii-Moriya interaction in the Hamiltonian $\hat{H}_Q$ defined in Eq.~\eqref{eq:ham_transformed}] with the expression Eq.~\eqref{eq:ASW} 
    from linear spin-wave theory. Model parameters are $\Delta = 0.5$, $D=0.5$, $\theta = \pi/2$
    and $\cos(Q) = \Delta + D[1-1/(2S)]$. Spin quantum numbers are $S=1,2$ and 3 in panels (a), (b) and (c), respectively. The power series for Eq.~\eqref{eq:onepoint} to second and fourth order are shown as dashed lines.
    }
    \label{fig:SW}
\end{figure}

Equation~\eqref{eq:ASW} in the main text describes the time evolution of the spin-helix amplitude $A$ in the linear spin-wave approximation for wave numbers satisfying $\cos(Q) = \Delta + D[1-1/(2S)]$. Figure~\ref{fig:SW} compares this expression with iTEBD simulations for the same parameters. Clearly, Eq.~\eqref{eq:ASW} does not accurately reproduce the numerical results, which is expected, since the linear spin-wave theory is valid for large $S$. When increasing the spin to $S=3$, we  observe some qualitative agreement. A major reason for the deviation at smaller spins appears to be the Dzyaloshinskii-Moriya interaction [see Eq.~\eqref{eq:ham_transformed}] that is neglected in the linear spin-wave theory. If this term is not included in the numerics the spin-wave result is reproduced more accurately.

Also shown in Fig.~\ref{fig:SW} is the time evolution of $A$ according to the power series expansion of Eq.~\eqref{eq:onepoint}, which quickly deviates from the numerical results for $St \gtrsim 0.5$. The second order term is given in Eq.~\eqref{eq:decay_2ndO}. For the parameters in Fig.~\ref{fig:SW} (a), (b) and (c) the fourth order coefficients are $\frac{91}{512}$, $\frac{13083}{4096}$ and $\frac{92645}{6912}$, respectively. 
The general expression is too lengthy to display here. 
It simplifies to zero for any $S$, however, if $D=0$ and the phantom condition is met. \\

\FloatBarrier


%

\end{document}